# Adsorbing Boundaries Stratify the Kinetics of Confined Colloidal Phase Separation

Kui Lin*, Haiming Lu

School of Civil Engineering, Sun Yat-Sen University, Guangzhou, Guangdong 510275, China

*Corresponding author: Kui Lin. E-mail: linkui66@mail.sysu.edu.cn

**Abstract:** Phase separation in network-forming materials couples mass transport to mechanical relaxation, yet how adsorbing boundaries reshape this coupling remains unclear. We compare fluid-particle-dynamics simulations resolving solvent hydrodynamics with free-draining Brownian dynamics for attractive colloids confined between parallel walls. Adsorbing boundaries do not slow coarsening uniformly but stratify the network into spatially distinct kinetic regimes. With hydrodynamics, the self-similar central network exhibits growth consistent with poroelastic relaxation, $L_{\mathrm{b}} \sim t^{1/2}$, whereas the adsorbed region follows an effective $L_{\mathrm{w}} \sim t^{1/3}$ law. In free-draining dynamics, central coarsening approaches $t^{1/3}$, while the boundary exponent remains time dependent over the accessible times. Direction-resolved displacements reveal suppressed wall-normal motion and reduced lateral mobility near the walls. Increasing wall attraction connects adsorbed islands into an extended layer while leaving local dense-phase packing nearly unchanged. For island-like domains, we propose a testable encounter-controlled scaling framework linking boundary growth to domain geometry, size-dependent mobility, and evolving adsorbed mass. Locally self-similar central coarsening thus coexists with distinct or nonstationary boundary kinetics, precluding a common dynamic-scaling length for the central and adsorbed regions over the accessible times; in both regions, the growth law depends on whether solvent hydrodynamics is resolved.

Coarsening is commonly understood as a progressive loss of spatial complexity. Following a quench, domains grow while preserving their statistical form, so that the evolving structure can be described by a single characteristic length. This dynamic-scaling picture underlies much of our understanding of phase ordering across fluids, alloys and soft materials [1,2]. It also implies a degree of kinetic uniformity: different parts of a system should share the same growth law, even when their local mobilities differ. Boundaries challenge this expectation. By breaking symmetry, restricting transport, and coupling selectively to one of the separating phases, a surface may do more than change the overall rate of coarsening: it may cause different regions of the same material to evolve according to different kinetic laws.

This possibility is particularly consequential in dynamically asymmetric mixtures, where phase separation creates a mechanically coherent network. If the dense phase relaxes more slowly than the deformation imposed by demixing, it can sustain stress and remain connected even at a minority volume fraction [3-7]. In deeply quenched attractive colloidal suspensions, contraction of this network requires solvent to move through the pores of the particle-rich skeleton. Structural evolution is therefore coupled to stress relaxation and collective solvent transport. In unconfined systems, this poroelastic mechanism produces growth consistent with $L \sim t^{1/2}$, whereas suppressing hydrodynamic coupling leads to slower, diffusion-like growth approaching $L \sim t^{1/3}$ [8,9]. The coarsening exponent thus reflects the mechanism by which the network relaxes, rather than simply the microscopic mobility of its constituent particles.

A solid surface can alter this mechanism in several ways [10-13]. No-slip confinement modifies solvent flow and many-body hydrodynamic coupling, while an attractive wall captures the dense phase and suppresses motion normal to the surface. The adsorbed network is consequently forced to reorganize within an effectively lower-dimensional environment while remaining mechanically coupled to the channel interior. Preferential surfaces are already known to direct phase separation through enrichment layers and composition waves [14,15], and hydrodynamic interactions remain important under no-slip confinement [16]. Much less is known, however, about the subsequent evolution of an adsorbed phase that is itself a connected, stress-bearing network. The central question is whether such a boundary merely slows an otherwise uniform coarsening process or creates a dynamically distinct boundary state with its own rate-limiting relaxation pathway.

Here we show that attractive confinement separates a phase-separating colloidal network into spatial regions governed by distinct kinetics. We compare fluid-particle-dynamics (FPD) simulations, which resolve incompressible solvent flow and many-body hydrodynamic coupling, with free-draining Brownian dynamics (BD), in which particles exchange momentum independently with an implicit background. This comparison allows us to distinguish the effects of collective solvent transport from constraints imposed directly by adsorption. In the hydrodynamic system, the channel interior retains three-dimensional self-similar growth consistent with poroelastic $t^{1/2}$ relaxation, whereas the adsorbed network coarsens more slowly, approximately following a boundary-constrained $t^{1/3}$ law. Without hydrodynamic coupling, the interior approaches the expected free-draining $t^{1/3}$ behaviour, but the adsorbed region slows further and does not reach a stationary scaling regime over the accessible times.

The resulting separation of length scales cannot be represented by a uniform reduction of particle mobility. As coarsening proceeds, the characteristic lengths of the wall and central regions progressively diverge, even though local self-similarity persists in the channel interior. Increasing the wall attraction strengthens this separation by transforming the adsorbed morphology from disconnected domains into a laterally extended layer and suppressing wall-normal rearrangements, while leaving the local packing of the dense phase largely unchanged. The boundary therefore acts primarily on mesoscale connectivity and transport rather than on the microscopic structure of the condensed phase. These findings establish adsorbing boundaries as selectors of coarsening mechanisms, rather than passive modifiers of kinetic prefactors. More generally, they show that local dynamic scaling can survive even when no single growing length describes a confined system as a whole. Such boundary-induced kinetic stratification should be relevant whenever phase separation couples transport to mechanical relaxation.

**Confined geometry and adsorption-selected morphology**

We consider $N$ colloids confined between parallel walls at $z = 0$ and $z = H$, with periodic boundary conditions in the $x$ and $y$ directions (Fig. 1a). The overall colloid volume fraction is $\phi = 0.10$, selected to produce a space-spanning colloid-rich network rather than isolated clusters. The channel heights are $H = 128$, 192 and 256 in simulation-grid units, and the colloid diameter is $a = 5.4$ in the same units.

Colloid pairs interact through a truncated Lennard–Jones potential,

$$U_{cc}(r) = 4\epsilon\left[\left(\frac{a}{r}\right)^{12} - \left(\frac{a}{r}\right)^{6}\right], r < r_c, \tag{1}$$

where $r$ is the centre-to-centre separation, $\epsilon$ is the colloid–colloid attraction energy and $r_c = 3a$ is the cutoff distance. The reduced temperature is $T^* = \frac{k_B T}{\epsilon} = 0.3$, corresponding to a deep quench in which interparticle attraction dominates thermal motion. Selected simulations are also performed at $T^* = 0$, representing a deterministic deep-quench limit, and are analysed only in terms of the Voronoi local-volume-fraction distributions and the particle and fluid velocity distributions.

The interaction between a colloid and either wall is represented by an integrated Lennard–Jones 9–3 potential,

$$U_w(z) = \epsilon_w \left[ \frac{2}{15} \left( \frac{\sigma_w}{z} \right)^9 - \left( \frac{\sigma_w}{z} \right)^3 \right], \tag{2}$$

where $z$ is the distance from the relevant wall, $\epsilon_w$ is the wall-attraction parameter and $\sigma_w = a$ sets its spatial range. The wall potential is truncated at $r_w = 5a$. We compare $\epsilon_w = 3\epsilon$ and $7\epsilon$, hereafter referred to as moderately and strongly adsorbing walls. The depths of the corresponding wall potentials exceed $k_B T$ by more than an order of magnitude. We do not call these states partial and complete wetting because contact angles and interfacial free energies have not been measured.

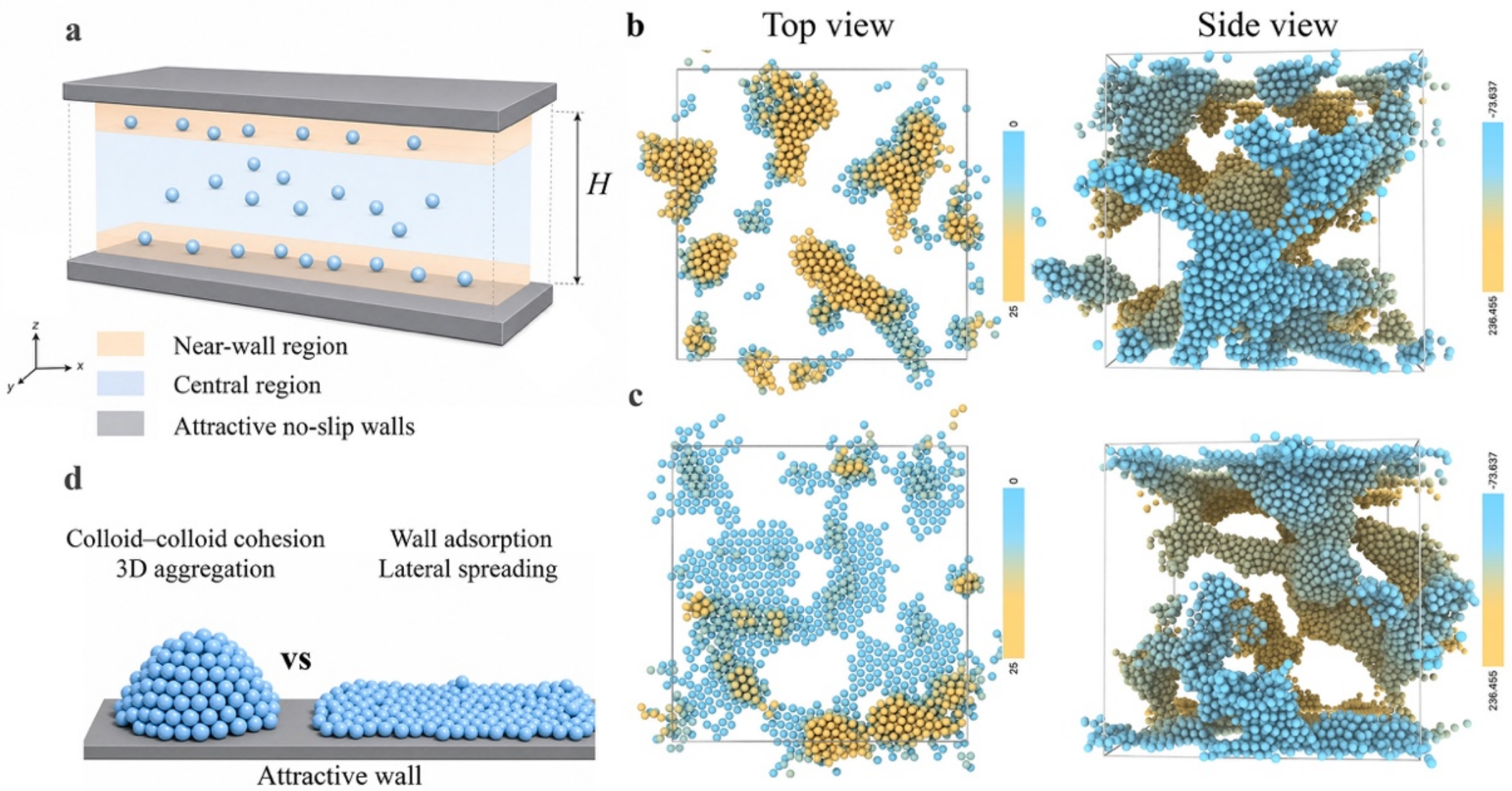

**Figure 1 | Attractive boundaries select distinct surface morphologies. a,** Simulation geometry. The colloidal suspension ($\phi = 0.10$, particle diameter $a = 5.4$) is periodic in $x$ and $y$ and confined by no-slip walls at $z = 0$ and $z = H$, with $H = 128$, 192 and 256. The near-wall and central regions used for most region-resolved analyses are indicated; the exact coarsening and three-dimensional structure-factor regions are specified in Methods. FPD resolves an explicit incompressible solvent, whereas BD couples each particle independently to an implicit background. **b,** Side and top views of FPD configurations at representative times for $\epsilon_w = 3\epsilon$. A connected central network coexists with discontinuous adsorbed islands having finite wall-normal thickness. **c,** Corresponding configurations for $\epsilon_w = 7\epsilon$. Strong adsorption produces a laterally extended, layer-like morphology occupying the first one or two particle layers. **d,** Schematic competition between colloid–colloid cohesion and wall adsorption. Three-dimensional aggregation maximizes colloidal coordination, whereas lateral spreading increases wall binding.

Immediately after the quench, the colloid-rich phase develops into a connected network in the central region. Near the walls, the morphology depends strongly on $\epsilon_w$. At $\epsilon_w = 3\epsilon$, adsorbed particles form discontinuous islands with appreciable wall-normal thickness (Fig. 1b). At $\epsilon_w = 7\epsilon$, they spread into a more laterally extended, layer-like structure (Fig. 1c). This morphological crossover reflects competition between colloid–colloid cohesion and wall adsorption. Under moderate attraction, complete lateral spreading would reduce the number of three-dimensional colloidal neighbours and hence incur a cohesive-energy penalty. Stronger wall attraction compensates for this loss and favours occupation of the first one or two adsorbed layers. The wall therefore selects how the dense phase partitions between lateral spreading and wall-normal stacking (Fig. 1d).

**Spatially resolved coarsening**

Global confinement breaks isotropy, so the wall and central regions are analysed separately. For the projected coarsening analysis, the near-wall regions comprise the first three density-peak layers adjacent to each boundary, whereas the central region spans $0.16H < z < 0.84H$. Other region-resolved analyses use the $4a$-thick wall and central slabs defined in Methods and Fig. 1a.

For each region, the particle coordinates are projected onto the $xy$ plane, and each particle is represented by a uniform circular disk of radius $a/2$, to construct the local areal-density field $\rho_{2D}(x, y, t)$. Its fluctuation is $\delta\rho_{2D}(x, y, t) = \rho_{2D}(x, y, t) - \langle\rho_{2D}(t)\rangle$. The corresponding in-plane structure factor is

$$S_{2D}(\mathbf{q}_{\parallel}, t) = | \mathcal{F}_{2D}\{\delta\rho_{2D}(x, y, t)\} |^2, \tag{3}$$

where $\mathcal{F}_{2D}$ denotes a two-dimensional Fourier transform and $\mathbf{q}_{\parallel} = (q_x, q_y)$. Angular averaging over wavevectors with magnitude $q_{\parallel} = (q_x^2 + q_y^2)^{1/2}$ gives $S(q_{\parallel}, t)$.

The characteristic in-plane wavevector is defined by its first moment,

$$q_{c,\parallel}(t) = \frac{\sum_{q_{\min}}^{q_{\max}} q_{\parallel} S(q_{\parallel}, t)}{\sum_{q_{\min}}^{q_{\max}} S(q_{\parallel}, t)}, \tag{4}$$

where the zero mode and the high-$q$ region dominated by the single-particle form factor are excluded. We denote the corresponding central and wall lengths by

$$L_{b,\parallel}(t) = \frac{2\pi}{q_{c,\parallel}^{(b)}(t)}, L_w(t) = \frac{2\pi}{q_{c,\parallel}^{(w)}(t)}. \tag{5}$$

Effective growth exponents are obtained from $q_{c,\parallel} \sim t^{-\nu}$. The main comparison uses the common conservative interval $15.8 < t/\tau_d \leq 30.6$, applied identically to all regions and both dynamical models. Because this interval spans only 0.29 decades, the resulting values are interpreted as intermediate-time effective exponents rather than asymptotic exponents. Wider model-specific intervals are used as robustness tests (Methods).

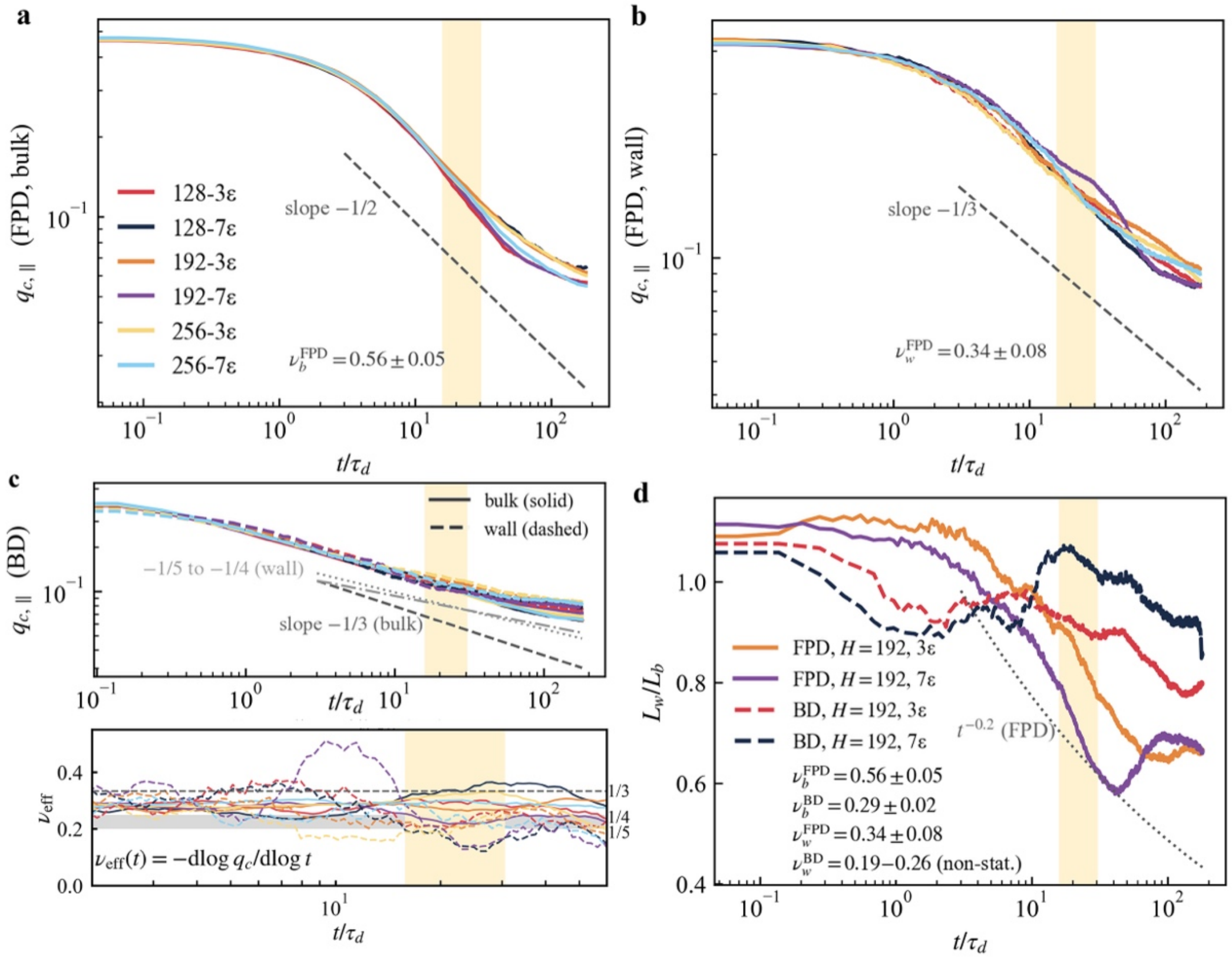

**Figure 2 | Adsorbed and central regions exhibit distinct coarsening kinetics. a,** In-plane characteristic wavevector $q_{c,\parallel}(t)$ of the FPD central region for all channel heights and wall attractions. The dashed line has slope $-1/2$; shading marks the common conservative fitting interval. **b,** Corresponding FPD near-wall wavevectors, with a $-1/3$ reference line. Upper and lower walls show statistically equivalent behaviour. **c,** Free-draining BD results. The central curves are compared with a $-1/3$ reference slope. Grey $-1/4$ and $-1/5$ lines near the wall are guides to idealized limiting behaviours and are not fitted exponents. The inset shows the running exponent $\nu_{\mathrm{eff}}(t) = -\frac{d\log q_{c,\parallel}}{d\log t}$, which has a stationary plateau in the central region but continues to evolve near the walls. **d,** Ratio $L_w(t)/L_{b,\parallel}(t)$, which decreases over the analysed interval. Effective exponents in the common interval are $\nu_b^{\mathrm{FPD}} = 0.56 \pm 0.05$, $\nu_w^{\mathrm{FPD}} = 0.34 \pm 0.08$ and $\nu_b^{\mathrm{BD}} = 0.29 \pm 0.02$. No unique exponent is assigned to the BD adsorbed region; its window-averaged values span approximately 0.19–0.26.

In the hydrodynamic FPD simulations, the projected central-region wavevector follows approximately $q_{c,\parallel}^{(b)} \sim t^{-1/2}$, giving $\nu_b^{\mathrm{FPD}} = 0.56 \pm 0.05$ across the investigated confinement conditions (Fig. 2a). The result is consistent, within the variation across physical conditions and fitting-window definitions, with the poroelastic prediction $\nu = 1/2$. The upper and lower wall curves decrease more slowly, with $\nu_w^{\mathrm{FPD}} = 0.34 \pm 0.08$, consistent with an effective $1/3$ boundary-constrained growth law (Fig. 2b). The separation between centre and wall is observed for both wall attractions and all investigated channel heights. Residual variations with adsorption strength are weak relative to the variation across confinement conditions and are treated as observational tendencies rather than distinct kinetic classes (Extended Data Fig. 2).

Free-draining BD exhibits a different form of stratification. The central region has the cleanest power-law regime in the data set, with $\nu_b^{\mathrm{BD}} = 0.29 \pm 0.02$. The value rises towards approximately 0.31 in the largest channel, approaching the free-draining $1/3$ prediction. By contrast, the BD adsorbed region does not exhibit a stationary exponent. Its window-averaged effective exponent varies from approximately 0.19 to 0.26 as the observation interval is changed, and its running exponent continues to evolve over the accessible time range (Fig. 2c and Extended Data Fig. 3). We therefore do not assign a unique asymptotic growth exponent to the BD wall region.

The wall-induced slowing cannot be attributed exclusively to modified hydrodynamic interactions because it is present in free-draining BD. Adsorption and the associated restriction of configurational motion introduce a slow boundary process even without nonlocal momentum transport. At the same time, FPD remains faster than BD over the common comparison interval, including near the wall, showing that collective solvent transport continues to influence boundary dynamics. The difference between the central and wall curves is not merely a constant time delay. A spatially uniform reduction of mobility would change the prefactor of the growth law while preserving its logarithmic slope. Instead, the ratio $L_w(t)/L_{b,\parallel}(t)$ decreases over the analysed interval (Fig. 2d). In FPD, the fitted exponents imply an approximate decrease $L_w/L_{b,\parallel} \sim t^{-0.2}$. In BD, the same ratio also decreases, but no single power is assigned because the wall exponent is non-stationary. At least two structural lengths are therefore required to describe the confined film. The system is not a homogeneous bulk coarsening process with a wall-dependent microscopic time scale; it contains spatially distinct kinetic regions.

**Three-dimensional self-similarity and transport-limited relaxation in the channel interior**

As time increases, the maximum of $S_b(q,t)$ shifts continuously towards smaller $q$, demonstrating growth of the central network scale (Fig. 3a). For conditions possessing an admissible self-similar interval, the scaled structure factors collapse according to $S_b(q,t) = L_b^3(t)F_b[qL_b(t)]$ (Fig. 3b and Extended Data Fig.4). Comparable collapse quality is found in FPD and BD, showing that self-similar central coarsening is not restricted to the hydrodynamic model. The smallest channels show earlier finite-size deviations and are treated as sensitivity cases rather than as the primary evidence for three-dimensional scaling.

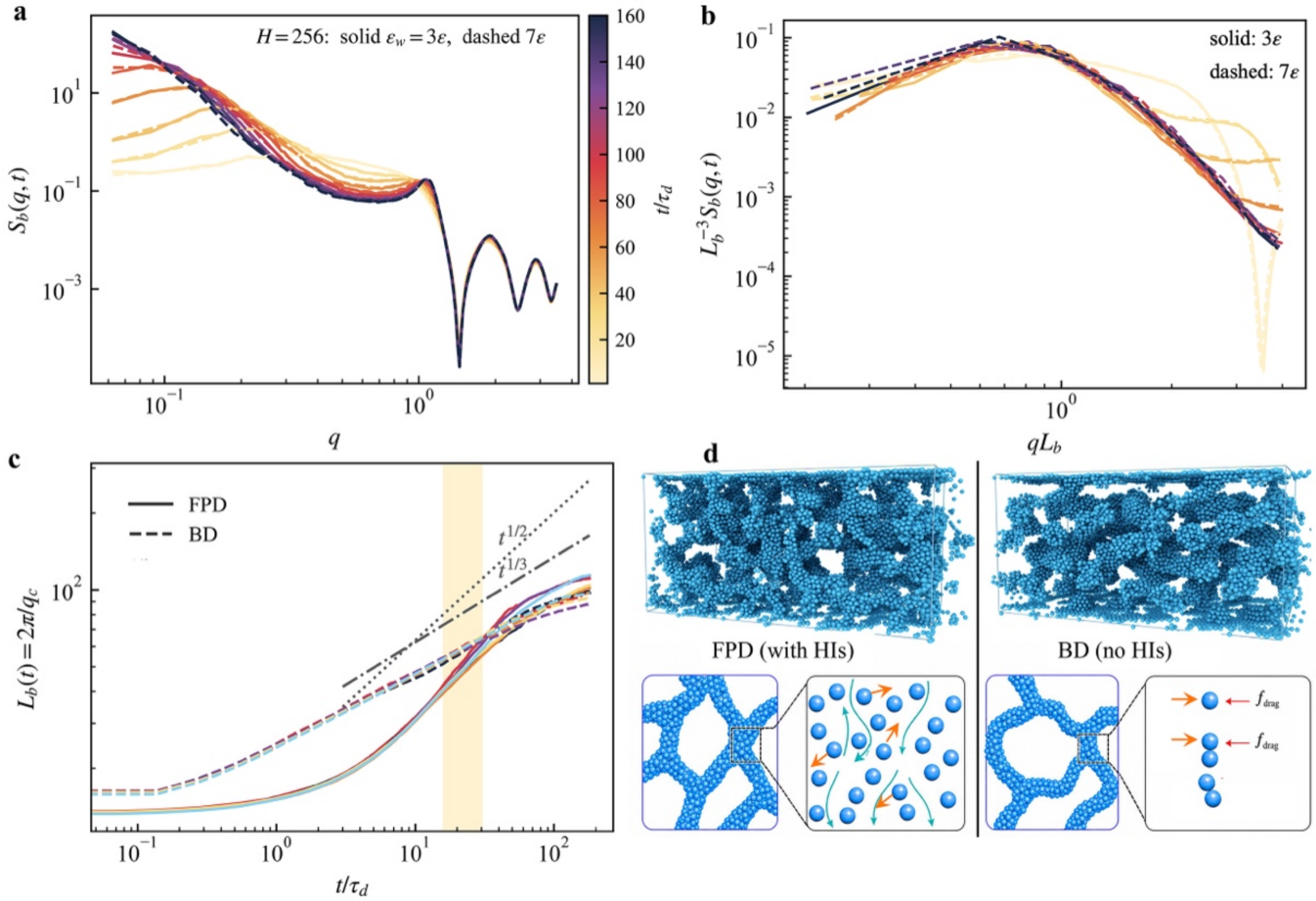

**Figure 3 | The central network retains three-dimensional self-similar coarsening. a,** Three-dimensional central structure factors $S_b(q,t)$ at increasing times. The dominant peak shifts towards lower $q$. **b,** Dynamic-scaling representation $L_b^{-3}S_b(q,t)$ versus $qL_b$. FPD and BD data collapse approximately over their admissible self-similar intervals; complete results are provided in Extended Data Fig.4. **c,** Central characteristic length $L_b(t)$, with $t^{1/2}$ and $t^{1/3}$ reference slopes for FPD and BD, respectively. **d,** Schematic comparison of the dominant relaxation pathways. In FPD, deformation of the colloid-rich skeleton requires relative solvent permeation, giving a poroelastic relaxation time $\tau_\mathrm{p} \sim L_\mathrm{b}^2/D_\mathrm{p}$ and the reference scaling $L_\mathrm{b} \sim t^{1/2}$. In free-draining BD, capillary driving is opposed by local particle friction, giving $\xi_0\,\mathrm{d}L/\mathrm{d}t \sim \gamma/L^2$ and $L \sim t^{1/3}$, where $\xi_0 = n\zeta_0$. Orange arrows denote the velocity of the colloidal skeleton, $\boldsymbol{v}_\mathrm{c}$; cyan arrows denote the relative solvent velocity, $\boldsymbol{v}_\mathrm{s} - \boldsymbol{v}_\mathrm{c}$; and red arrows denote local BD drag. The relations shown are scaling descriptions rather than quantitative predictions of the prefactors.

The central FPD growth law is consistent with poroelastic relaxation of the stress-bearing colloid-rich network [8,17]. Deformation of this network requires solvent motion relative to the colloidal skeleton, such that relaxation over the characteristic distance $L_\mathrm{b}$ is controlled by solvent permeation through the dense phase. Provided that the dense-phase permeability and longitudinal modulus vary only weakly during the scaling regime, the poroelastic relaxation time satisfies

$$\tau_\mathrm{p}(L_\mathrm{b}) \sim \frac{L_\mathrm{b}^2}{D_\mathrm{p}},\ D_\mathrm{p} \sim \frac{kM}{\eta_\mathrm{s}}, \tag{6}$$

where $k$ is the permeability of the colloid-rich phase, $\eta_\mathrm{s}$ is the solvent viscosity and $M$ is its effective longitudinal modulus. For an isotropic linear-elastic skeleton, $M = K + 4G/3$, where $K$ and $G$ are the bulk and shear moduli, respectively. Identifying the poroelastic relaxation time with the elapsed coarsening time then gives

$$L_\mathrm{b}(t) \sim \left(D_\mathrm{p}t\right)^{1/2}. \tag{7}$$

The free-draining BD model has a different dissipation pathway. If $\zeta_0$ is the single-particle drag coefficient, the corresponding friction coefficient per unit volume is $\xi_0 = n\zeta_0$, where $n$ is the particle number density. Balancing the capillary driving-force density, of order $\gamma/L^2$, against local friction gives

$$\xi_0 \frac{\mathrm{d}L}{\mathrm{d}t} \sim \frac{\gamma}{L^2}, \tag{8}$$

and hence

$$L(t) \sim \left(\frac{\gamma t}{\xi_0}\right)^{1/3}. \tag{9}$$

The central-region exponents extracted from these growth curves approach the poroelastic prediction $\alpha = 1/2$ in FPD and the free-draining prediction $\alpha = 1/3$ in BD as the channel height increases (Fig. 3c). The central regions therefore reproduce the established distinction between poroelastic and free-draining network relaxation [8,9]. These two relaxation pathways are summarized schematically in Fig. 3d. The attraction supplies the thermodynamic driving force and mechanical stress, but the growth exponent follows from the scale dependence of the rate-limiting relaxation time, not from the Lennard–Jones potential alone.

**Boundary-constrained coarsening**

The adsorbed regions do not follow the central laws of their respective models. Adsorption localizes particles in $z$, changes the local topology and introduces momentum transfer to the solid boundary. Their wall-parallel force balance can be represented schematically by

$$\nabla_{\parallel} \cdot \boldsymbol{\sigma}_{\parallel} - \xi_w(L_w)\mathbf{u}_{c,\parallel} + \Gamma_{\parallel}(\mathbf{u}_{s,\parallel} - \mathbf{u}_{c,\parallel}) \simeq 0, \tag{10}$$

where $\boldsymbol{\sigma}_{\parallel}$ is the in-plane stress, $\mathbf{u}_{c,\parallel}$ and $\mathbf{u}_{s,\parallel}$ are the in-plane colloid and solvent velocities, $\Gamma_{\parallel}$ is their effective mutual friction and $\xi_w$ is an effective wall-associated resistance. This resistance includes direct adsorption, crowding, substrate-coupled hydrodynamic dissipation and structural rearrangement within the adsorbed layer.

If an interfacial or mechanical stress of scale $\gamma_w/L_w$ varies over a distance $L_w$, its divergence scales as $\gamma_w/L_w^2$. If the wall-associated resistance dominates and is approximately independent of scale,

$$\frac{\mathrm{d}L_w}{\mathrm{d}t} \sim \frac{\gamma_w}{\xi_w L_w^2}, \tag{11}$$

which gives $L_w \sim t^{1/3}$. More generally, if

$$\xi_w(L_w) \sim L_w^m, \tag{12}$$

then

$$L_w(t) \sim t^{1/(3+m)}. \tag{13}$$

The FPD wall exponent, $\nu_w^{\mathrm{FPD}} = 0.34 \pm 0.08$, is consistent with the approximately scale-independent limit $m \simeq 0$.

The BD adsorbed region behaves differently. Its running exponent has no stationary plateau, and the corresponding window-averaged values span approximately 0.19–0.26. This indicates an effective resistance whose scale dependence evolves with time. We therefore do not infer a precise value of $m$ from the BD data. Two observations suggest possible origins of this additional resistance. First, collective solvent transport can redistribute material within the adsorbed region without requiring every particle to move independently against its neighbours; this channel is absent in BD by construction. Second, BD exhibits the strongest medium-range positional order, particularly near the wall (Fig. 5d). Increased local ordering may hinder cooperative rearrangement and generate progressively slower boundary kinetics. The present data suggest, but do not establish, this ordering–mobility connection.

At $\epsilon_w = 3\epsilon$, lateral evolution can involve translation, deformation and merger of finite adsorbed islands. At $\epsilon_w = 7\epsilon$, the relevant events are more likely continuous restructuring of a connected layer and elimination of low-density holes. The available observables do not distinguish these elementary processes, so both states are conservatively described as boundary-constrained in-plane restructuring.

**Adsorption-induced mobility anisotropy**

The microscopic dynamical consequences of confinement are quantified by separating particle displacements into components parallel and perpendicular to the walls. For lag time $\tau$,

$$\mathrm{MSD}_{\parallel}(\tau) = \langle [x(t+\tau) - x(t)]^2 + [y(t+\tau) - y(t)]^2 \rangle, \tag{14}$$

whereas

$$\mathrm{MSD}_{\perp}(\tau) = \langle [z(t+\tau) - z(t)]^2 \rangle. \tag{15}$$

Because the phase-separating structure is ageing and non-stationary, these quantities are finite-time measures of particle motion and should not automatically be interpreted as equilibrium diffusion coefficients. Direction-resolved displacements for all parameter combinations are reported in Fig. 4 and Extended Data Fig. 5. For all investigated conditions, particles in the central region have larger in-plane displacements than particles near the wall (Fig. 4a). Adsorption therefore inhibits not only wall-normal motion but also the lateral rearrangements required for coarsening. Wall-normal suppression is substantially stronger. In the adsorbed regions, $\mathrm{MSD}_{\perp} \ll \mathrm{MSD}_{\parallel}$, indicating that particles are confined to a narrow range of $z$ (Fig. 4b). This anisotropy is consistent with the distinction between three-dimensional central relaxation and boundary-constrained in-plane evolution.

At fixed $H$, the central FPD simulations show larger wall-normal displacements for $\epsilon_w = 7\epsilon$ than for $3\epsilon$(Fig. 4c). The corresponding trend is weak or absent in BD. Because the measured displacement contains directed adsorption, network contraction and advection in addition to fluctuations, this difference should not be described as enhanced equilibrium diffusion. It is instead consistent with stronger adsorption generating collective solvent and network motion that is transmitted into the channel interior. The dependence on $H$ also differs between FPD and BD. Central BD curves remain comparatively close for different wall attractions and channel heights, whereas the FPD curves separate systematically (Fig. 4d). In BD, particles exchange momentum independently with a featureless background. In FPD, adsorption and network contraction perturb a shared incompressible solvent field, allowing boundary conditions to influence motion beyond the direct range of the wall potential.

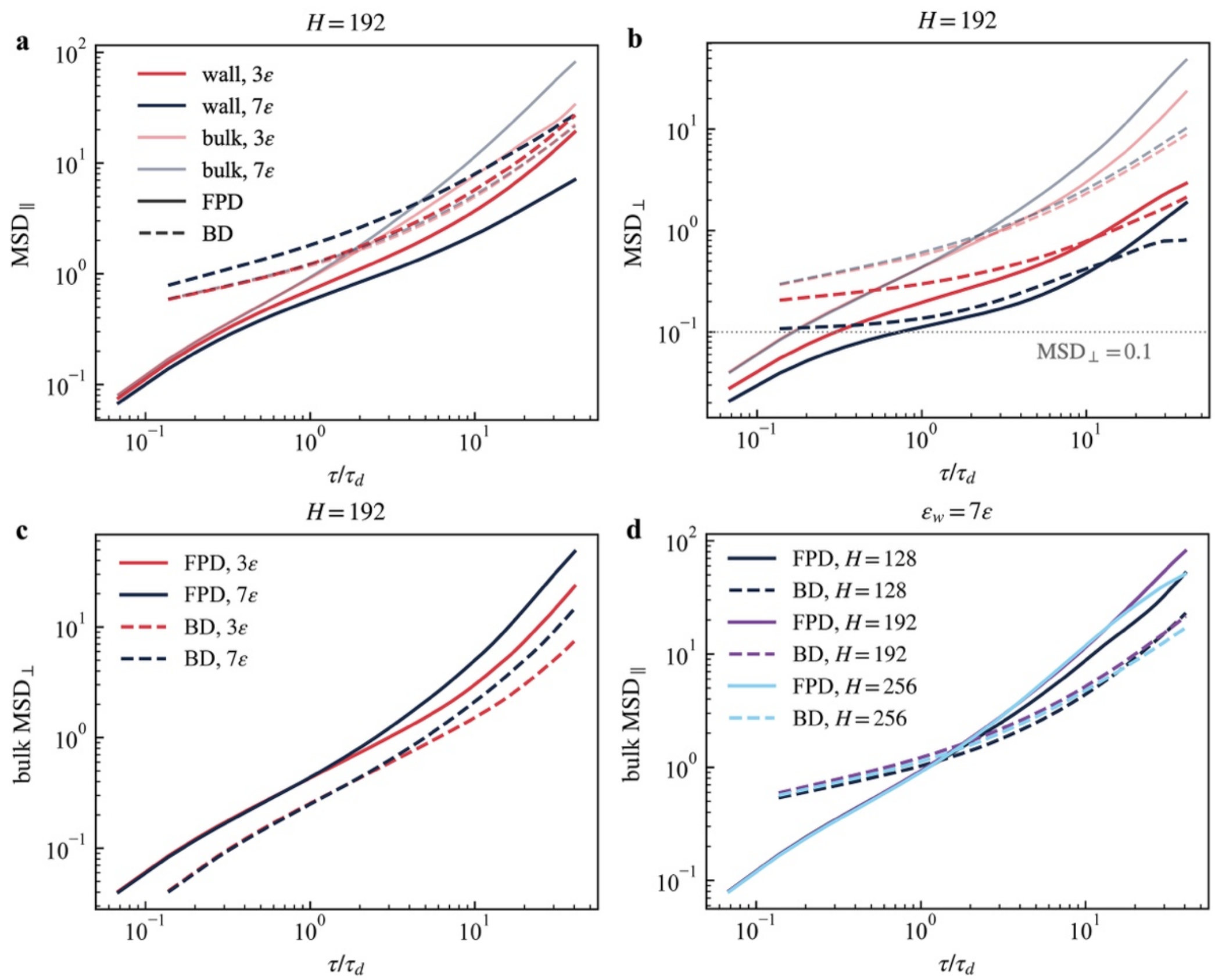

**Figure 4 | Wall adsorption produces anisotropic particle motion. a,** In-plane mean-squared displacement $\mathrm{MSD}_{\parallel}(\tau)$in central and near-wall regions. Adsorbed particles move more slowly parallel to the wall. **b,** Wall-normal displacement $\mathrm{MSD}_{\perp}(\tau)$. Near-wall normal motion is strongly suppressed relative to the central region and to the in-plane component. **c,** Dependence on wall attraction at fixed $H$. Central FPD wall-normal motion is larger for $\epsilon_w = 7\epsilon$ than for $3\epsilon$, whereas the corresponding BD curves remain comparatively close. **d,** Dependence on channel height at strong adsorption. FPD displacements vary more strongly with $H$than BD displacements, consistent with nonlocal transmission of boundary perturbations through the solvent.

### Hydrodynamic interactions and the hierarchy of relaxation pathways

The FPD–BD comparison provides a unified interpretation of the coarsening and displacement statistics. In the central region, FPD retains solvent pressure, backflow and momentum conservation and produces poroelastic growth approaching $t^{1/2}$. BD replaces these collective processes with diagonal particle friction and gives slower central growth approaching $t^{1/3}$. The two models stratify in different ways. In FPD, the central and adsorbed regions exhibit distinct but approximately stationary effective laws, with $\nu_b^{\mathrm{FPD}} \simeq 0.56$ and $\nu_w^{\mathrm{FPD}} \simeq 0.34$. In BD, only the central region has a stable power-law regime; the adsorbed region displays a continuously evolving effective exponent. Kinetic stratification is therefore common to both models, but appears as coexistence of two slow laws in FPD and as coexistence of bulk scaling and boundary ageing in BD.

The presence of wall-induced slowing in both models shows that adsorption introduces a kinetic bottleneck that is not exclusively hydrodynamic. Conversely, the persistence of faster FPD dynamics over the common comparison interval shows that the wall does not eliminate collective solvent transport. The physical presence of hydrodynamic interactions and their role in selecting the rate-limiting process are distinct questions. The observed relaxation time may be represented schematically as $\tau_{\text{obs}} = \max(\tau_{\text{poro}}, \tau_{\text{FD}}, \tau_{\text{ads}}, \tau_{\parallel})$, where $\tau_{\text{poro}}$ is the poroelastic drainage time, $\tau_{\text{FD}}$ is the free-draining mechanical-relaxation time, $\tau_{\text{ads}}$characterizes restructuring within the wall potential and $\tau_{\parallel}$is the time required for collective in-plane reorganization.

In the central FPD region, $\tau_{\text{obs}}$ is controlled primarily by poroelastic relaxation. In central BD, it is controlled by free-draining mechanical relaxation. Near an attractive wall, adsorption-constrained and in-plane restructuring times become dominant. The maximum form identifies the leading contribution rather than implying that the remaining processes are absent. The boundary therefore changes which process is rate limiting; it does not simply screen hydrodynamic interactions or multiply all microscopic time scales by a constant factor.

**Morphology changes more strongly than local packing**

Having established that the wall alters the relaxation pathway, we next ask how deeply it alters structure. Three measures—projected coverage, local density and pair correlations—give a consistent answer. The near-wall morphology is first characterized by an effective projected box-counting dimension. If $N_{\text{box}}(\ell)$ is the number of occupied boxes of side length $\ell$, then $N_{\text{box}}(\ell) \sim \ell^{-d_{\text{box}}}$ over an intermediate scale range. For the near-wall regions, particle positions are projected onto the $xy$plane and two-dimensional boxes are used. The island-like morphology at $\epsilon_w = 3\epsilon$ has $d_{\text{box}}^{2D} \simeq 1.5$, whereas the extended layer at $\epsilon_w = 7\epsilon$ has $d_{\text{box}}^{2D} \simeq 1.8$ (Fig. 5a). The latter approaches the plane-filling value of two. The central network is analysed using three-dimensional boxes and has $d_{\text{box}}^{3D} \simeq$ 2.0–2.3, with much weaker dependence on wall attraction (Fig. 5b). These values are finite-range morphological descriptors rather than universal fractal dimensions.

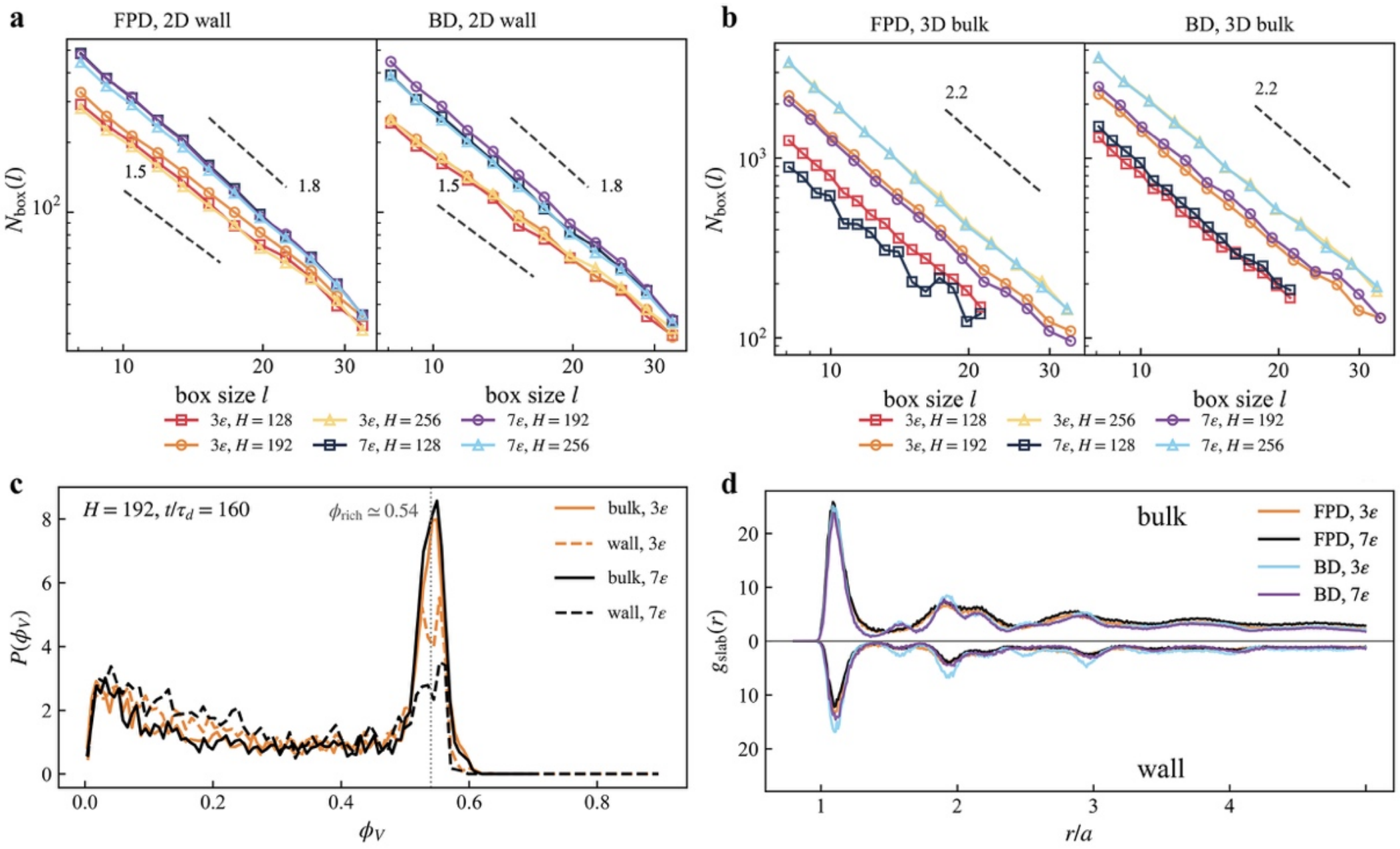

**Figure 5 | Adsorption reorganizes topology more strongly than dense-phase packing. a,** Two-dimensional box-counting analysis of projected wall structures. The island-like morphology gives $d_{\text{box}}^{2D} \simeq 1.5$, whereas the extended layer gives $d_{\text{box}}^{2D} \simeq 1.8$. **b,** Three-dimensional box-counting analysis of the central network, giving $d_{\text{box}}^{3D} \simeq$ 2.0–2.3. **c,** Probability distributions of the Voronoi local volume fraction. The colloid-rich maximum remains near $\phi_{\text{rich}} \simeq 0.54$ for different wall attractions and statistical regions. **d,** Slab-resolved radial distribution functions. BD exhibits pronounced secondary

structures in both central and near-wall regions; FPD decays more smoothly in the centre but develops sharper secondary features near the wall. These data indicate enhanced positional order but do not establish crystallization or a causal relationship to the slow wall kinetics.

Local packing is measured using the Voronoi volume fraction

$$\phi_{V,i} = \frac{V_p}{V_{\mathrm{Voro},i}} = \frac{\pi a^3/6}{V_{\mathrm{Voro},i}}, \tag{16}$$

where $V_{\mathrm{Voro},i}$ is the Voronoi-cell volume associated with particle $i$. The tessellation is periodic in $x$ and $y$, and near-wall cells are clipped by the physical boundary. The colloid-rich maximum of $P(\phi_V)$ remains close to $\phi_{\mathrm{rich}} \simeq 0.54$ for both wall attractions and for central and near-wall regions (Fig. 5c and Extended Data Fig. 6). This value is also close to the dense-phase volume fraction reported for unconfined network-forming colloidal phase separation [8]. The stability of the high-density maximum does not imply that wall and central structures are microscopically identical. Their topology, orientational order and pair correlations can differ substantially. It shows instead that adsorption changes where the dense phase is located and how it connects more strongly than it changes its characteristic local packing density.

Pair correlations refine this picture at intermediate range. To correct for shell truncation by the slab boundaries, the available area of a spherical shell centred at $z_i$ is $A_{\mathrm{valid}}(r, z_i) = 2\pi r h_{\mathrm{valid}}$, where

$$h_{\mathrm{valid}} = \max\,[0, \min\,(z_i + r, z_{\max}) - \max\,(z_i - r, z_{\min})]. \tag{17}$$

The slab radial distribution function is then

$$g_{\mathrm{slab}}(r) = \frac{H(r)}{\rho_{\mathrm{slab}} \Delta r \sum_i A_{\mathrm{valid}}\,(r, z_i)}, \tag{18}$$

where $H(r)$ is the pair-count histogram, $\Delta r = 0.05$ and $\rho_{\mathrm{slab}}$ is the mean number density in the slab. BD displays pronounced secondary features beyond the nearest-neighbour peak in both central and wall regions (Fig. 5d). Results for all parameter combinations are provided in Extended Data Fig. 7. In FPD, the central curves decay more smoothly, whereas sharper secondary features appear near the walls. These observations indicate enhanced medium-range positional order near the adsorbing boundaries, reminiscent of ordering pathways reported in viscoelastic phase separation [18], and stronger ordering in free-draining BD than in the hydrodynamic central network. The BD wall region, which has the strongest secondary structure, also exhibits the most strongly non-stationary coarsening. Nevertheless, a split secondary peak is insufficient to prove crystallization, and the present analysis cannot establish a causal relationship between ordering and kinetic slowing. Independent bond-orientational or common-neighbour analyses would be required. Taken together, the structural measures show that adsorption acts primarily at the mesoscale: it reorganizes coverage, connectivity and medium-range order while leaving the characteristic local packing density of the colloid-rich phase essentially unchanged.

**Discussion**

Our simulations reveal a spatial separation of mechanical relaxation in a confined, phase-separating colloidal network. The central region retains approximately self-similar three-dimensional coarsening, whereas the adsorbed region evolves through a slower and more constrained pathway.

In FPD, the central network exhibits growth consistent with poroelastic $t^{1/2}$ coarsening, whereas the adsorbed region exhibits a slower effective law close to $t^{1/3}$. In BD, the central region shows robust power-law growth approaching the free-draining $t^{1/3}$ limit, whereas the adsorbed region does not reach a stationary growth exponent over the accessible time range. Attractive confinement therefore produces two forms of kinetic stratification: coexistence of distinct effective scaling laws in the hydrodynamic system, and coexistence of bulk scaling with non-stationary boundary ageing in the free-draining system.

The central result is not simply that an attractive wall reduces particle mobility. A spatially uniform, scale-independent reduction of mobility would alter the kinetic prefactor while preserving the asymptotic growth exponent. Instead, the central and near-wall lengths evolve at different relative rates, such that $L_{\mathrm{w}}/L_{\mathrm{b},\parallel}$ changes with time. The confined film therefore cannot be described by a single global coarsening length.

Moderately and strongly adsorbing walls produce qualitatively different surface topologies. At $\epsilon_{\mathrm{w}} = 3\epsilon$, colloid–colloid cohesion remains competitive with wall adsorption, producing finite islands with appreciable wall-normal thickness. At $\epsilon_{\mathrm{w}} = 7\epsilon$, stronger wall binding favours lateral coverage and produces a laterally extended adsorbed layer approaching a plane-filling morphology over the resolved scales. These pronounced topological changes are accompanied by only modest changes in the measured wall-growth kinetics.

**A scaling framework for island-like surface coarsening**

The island-like morphology observed at moderate wall attraction suggests an encounter-controlled coarsening pathway. Diffusion and coalescence of mobile domains have long been used to describe aggregation and thin-film coarsening, including cases in which the diffusivity depends on domain mass [19,20]. Here we generalize this standard argument by allowing the surface domains to be non-compact and by treating their size and separation as independent scaling variables. Let $N_s$ denote the number of particles in a surface domain and $R_s$ its characteristic in-plane size. We write $N_s \sim R_s^{d_f}$, where $d_f$ is the mass–size exponent of an individual domain and is distinct from the projected box-counting dimension of the complete wall morphology.

If the in-plane domain diffusivity scales as $D_c(N_s) \sim N_s^{-\mu}$, and the mean domain separation scales as $\ell_s \sim R_s^{\zeta}$, then the encounter time is

$$\tau_{\mathrm{enc}} \sim \frac{\ell_s^2}{D_c} \sim R_s^{2\zeta+\mu d_f}. \tag{19}$$

Identifying this time with the elapsed coarsening time gives

$$R_s(t) \sim t^{1/(2\zeta+\mu d_f)}, \nu_w = \frac{1}{2\zeta+\mu d_f}. \tag{20}$$

For a closed adsorbed subsystem with approximately conserved surface mass,

$$n_s N_s \simeq \Sigma_w = \text{constant}, \tag{21}$$

where $n_s$ is the domain number per unit area and $\Sigma_w$ is the adsorbed particle number per unit wall area. Because $\ell_s \sim n_s^{-1/2}$, $\ell_s \sim R_s^{d_f/2}$, and therefore

$$\nu_w = \frac{1}{d_f(1+\mu)}. \tag{22}$$

If the adsorbed mass instead grows as $\Sigma_w(t) \sim t^{\alpha}$, a population-balance argument gives

$$\nu_w = \frac{1+\alpha}{d_f(1+\mu)}. \tag{23}$$

These relations provide a falsifiable mechanism rather than a parameter-free interpretation of the present data. The BD wall exponent is non-stationary, and neither $d_f$, $\mu$ nor $\alpha$ has been measured independently. We therefore do not invert the observed effective exponents to determine these quantities.

The decisive tests are direct measurements of the domain mass–size relation $N_s(R_s)$, domain mobility $D_c(N_s)$, domain separation $\ell_s(t)$ and surface-mass evolution $\Sigma_w(t)$. These observables would distinguish encounter-controlled coalescence from the alternative wall-resistance description, which produces the same FPD $1/3$ law when the effective boundary resistance is scale independent. The encounter picture is most directly applicable to the island-like state at $\epsilon_w = 3\epsilon$. In the connected state at $7\epsilon$, the characteristic in-plane length may instead measure the elimination of low-density holes or continuous collective restructuring of the adsorbed network.

**Broader implications**

The coexistence of central and near-wall kinetics demonstrates that global dynamic scaling can fail as a description of the complete confined film even when the central region remains locally self-similar. A confined phase-separating material may therefore require spatially resolved characteristic lengths rather than a single global domain scale. This framework should apply more broadly to network-forming soft materials adjacent to attractive, frictional or impermeable interfaces, including colloidal gels in microchannels, polymer-rich networks near solid substrates and stress-bearing biomolecular assemblies adjacent to membranes. In such systems, a boundary acts not only as a thermodynamic surface field but also as a selector of transport dimensionality and mechanical relaxation pathway.

The robust conclusion is boundary-induced kinetic stratification: self-similar bulk-like mechanical coarsening in the channel centre coexists with boundary-constrained slow scaling in FPD or non-stationary ageing in BD. At the same time, adsorption strongly reorganizes mesoscale topology while leaving dense-phase packing nearly unchanged. Boundaries therefore select how stress relaxes and material is transported, rather than merely slowing the entire phase-separating system uniformly.

## Methods

### Fluid-particle dynamics

In FPD, each colloid is represented by a highly viscous diffuse region embedded in a Newtonian solvent [21,22]. The viscosity field is

$$\eta(\mathbf{r}, t) = \eta_s + (\eta_c - \eta_s) \sum_{i=1}^{N} \psi_i(\mathbf{r}, t), \tag{24}$$

where $\psi_i$ is the profile function of particle $i$. We use $\eta_s = 1.17$ and $\eta_c = 50\eta_s$. The colloid and solvent densities are both set to $\rho = 1$.

The velocity field obeys incompressibility, $\nabla \cdot \mathbf{v} = 0$, and evolves according to

$$\rho\left(\frac{\partial \mathbf{v}}{\partial t} + \mathbf{v} \cdot \nabla \mathbf{v}\right) = -\nabla p + \nabla \cdot [\eta(\mathbf{r})(\nabla \mathbf{v} + \nabla \mathbf{v}^T)] + \mathbf{f}_p + \nabla \cdot \boldsymbol{\sigma}^R. \tag{25}$$

Here $p$ is the pressure, $\mathbf{f}_p$ is the force density generated by colloid–colloid and colloid–wall interactions, and $\boldsymbol{\sigma}^R$ is the random stress. Its covariance is

$$\begin{aligned}\langle \sigma_{ij}^R(\mathbf{r},t)\sigma_{kl}^R(\mathbf{r}',t')\rangle = \quad & 2k_B T\eta(\mathbf{r})(\delta_{ik}\delta_{jl} + \delta_{il}\delta_{jk}) \\ & \times \delta(\mathbf{r}-\mathbf{r}')\delta(t-t').\end{aligned} \tag{26}$$

No-slip conditions are imposed at both physical walls.

**Brownian dynamics**

The free-draining BD model obeys

$$\zeta_0 \frac{d\mathbf{R}_i}{dt} = \mathbf{F}_i^{cc} + \mathbf{F}_i^{w} + \mathbf{F}_i^{R}, \tag{27}$$

where $\zeta_0$ is the single-particle friction coefficient, $\mathbf{F}_i^{cc}$ and $\mathbf{F}_i^{w}$ are conservative colloid–colloid and wall forces, and $\mathbf{F}_i^{R}$ is Gaussian noise satisfying

$$\langle F_{i\alpha}^R(t)F_{j\beta}^R(t')\rangle = 2k_B T\zeta_0\delta_{ij}\delta_{\alpha\beta}\delta(t-t'). \tag{28}$$

BD retains an implicit solvent through friction and thermal noise but neglects non-diagonal many-body mobility and solvent momentum transport.

**Three-dimensional structural analysis of the channel interior**

To establish whether the interior preserves genuine three-dimensional network coarsening, we reconstruct a volumetric density field. Each colloid is represented by a finite-volume shape function $\Phi$,

$$\rho_{3D}(\mathbf{r},t) = \sum_{i=1}^{N_b} \Phi(|\,\mathbf{r} - \mathbf{r}_i\,|), \tag{29}$$

where $\mathbf{r}_i$ is the centre of particle $i$and $N_b$ is the number of particles in the interior analysis slab. Taking $R = a/2$ as the particle radius,

$$\Phi(r) = \begin{cases} \dfrac{6}{\pi a^3}, & r \le R, \\ 0, & r > R. \end{cases} \tag{30}$$

Because the extracted slab is not periodic in $z$, a Tukey window $W(z)$ is applied to suppress Fourier leakage from its boundaries. The three-dimensional structure factor is

$$S_b(q,t) = \langle \frac{|\,\mathcal{F}_{3D}\{\delta\rho_{3D}(\mathbf{r},t)W(z)\}\,|^2}{N_b M_p^2 W_{\text{loss}}} \rangle_{|\mathbf{q}|=q}, \tag{31}$$

where $\delta\rho_{3D} = \rho_{3D} - \langle\rho_{3D}\rangle$, $M_p$ is the reconstructed mass of one particle and $W_{\text{loss}}$ corrects for the spectral weight removed by windowing.

The characteristic three-dimensional wavevector is

$$q_b(t) = \frac{\sum_{q_{\min}}^{q_{\max}} q\, S_b(q,t)}{\sum_{q_{\min}}^{q_{\max}} S_b\,(q,t)}, \tag{32}$$

and the corresponding network scale is $L_b(t) = \frac{2\pi}{q_b(t)}$.

**Hydrodynamic validation**

The no-slip implementation is tested using pure-solvent Poiseuille flow. A constant body-force density $G = 10^{-4}$ is applied in the $x$ direction. The analytical steady velocity profile is

$$v_x(z) = \frac{G}{2\eta_s}(z - z_{\min})(z_{\max} - z), \tag{33}$$

where $z_{\min}$ and $z_{\max}$ are the effective hydrodynamic wall positions.

The numerical profile follows the analytical parabola and approaches zero at both walls (Extended Data Fig. 1). A free-wall fit gives $z_{\min} \simeq 2.2, z_{\max} \simeq 125.8, \eta_{\text{eff}} \simeq 1.18$, within approximately 1% of the input $\eta_s = 1.17$, with $R^2 = 0.9993$.

**Velocity statistics**

Particles belonging to the colloid-rich skeleton are selected using $\phi_V > 0.16$. The wall region is defined as the region within $4a$ of either boundary. The speed of particle $i$ is

$$V_i = (v_{x,i}^2 + v_{y,i}^2 + v_{z,i}^2)^{1/2}. \tag{34}$$

At each time, the particle speed distribution is scaled as $\langle V\rangle P(V)$ versus $V/\langle V\rangle$.

For FPD, the local fluid speed is

$$v(\mathbf{r}) = |\,\mathbf{v}(\mathbf{r})\,|, \tag{35}$$

and its distribution is scaled as $\bar{v}P(v)$ versus $v/\bar{v}$, where $\bar{v}$ is the regional mean speed. The central portions display approximate temporal collapse, whereas high-speed tails and wall distributions are more sensitive to intermittent restructuring and finite statistics (Extended Data Fig. 8). These distributions support approximate similarity of the dominant velocity population but do not establish a universal velocity distribution.

**Statistical analysis**

Characteristic wavevectors are calculated over fixed wavevector intervals. Effective growth exponents are initially extracted over a common conservative interval, $15.8 < t/\tau_d \leq 30.6$, chosen to enable direct comparisons across dynamical models and spatial regions while excluding the initial crossover and late-time finite-size effects. Because this interval spans only 0.29 decades, the resulting values are interpreted as intermediate-time effective exponents rather than asymptotic exponents. Their robustness is assessed using wider model-specific fitting intervals and, where applicable, intervals constrained by the onset of three-dimensional dynamic scaling (Extended Data Fig. 9). For the FPD central and wall regions and the BD central region, the resulting group-averaged exponents differ from those obtained over the common interval by no more than 0.007. By contrast, the estimated exponent for the BD wall region changes from approximately 0.19 to 0.26, and its running exponent exhibits no stationary plateau. We therefore do not assign a unique asymptotic growth exponent to the BD wall region.

Three-dimensional structure-factor collapse is used as an independent test of self-similar central coarsening. Curves for which the collapse onset does not overlap the conservative fitting interval are refitted over model-specific admissible intervals. The resulting exponent changes are 0.003–0.011 and do not alter the hierarchy of central and boundary dynamics. Quoted exponents are means over the available $H$–$\epsilon_w$ combinations, and quoted uncertainties are standard deviations over those curves. They quantify variation across physical conditions rather than the standard error of statistically identical repeats.

For the projected coarsening analysis, the wall regions comprise the first three density-peak layers adjacent to each wall, and the central region spans $0.16H < z < 0.84H$. The three-dimensional structure-factor analysis uses the interior slab $0.08H < z < 0.92H$, with a Tukey window applied in $z$. Displacement, box-counting, Voronoi and velocity analyses use the $4a$-thick wall regions and the corresponding central slab.

## Acknowledgements

This work was supported by the Science and Technology Program of Guangdong Province (Grant No. 2024QN11G117) and the start-up funding from Sun Yat-Sen University to Kui Lin.

**Author contributions**

**Kui Lin:** Conceptualization, Investigation, Methodology, Software, Formal analysis, Visualization, Validation, Resources, Supervision, Funding acquisition, Project administration, Writing – original draft, and Writing – review & editing. **Haiming Lu:** Investigation, Data curation, Formal analysis, Validation, and Writing – review & editing.

**Competing interests**

The authors declare no competing interests.

**Extended Data figure captions**

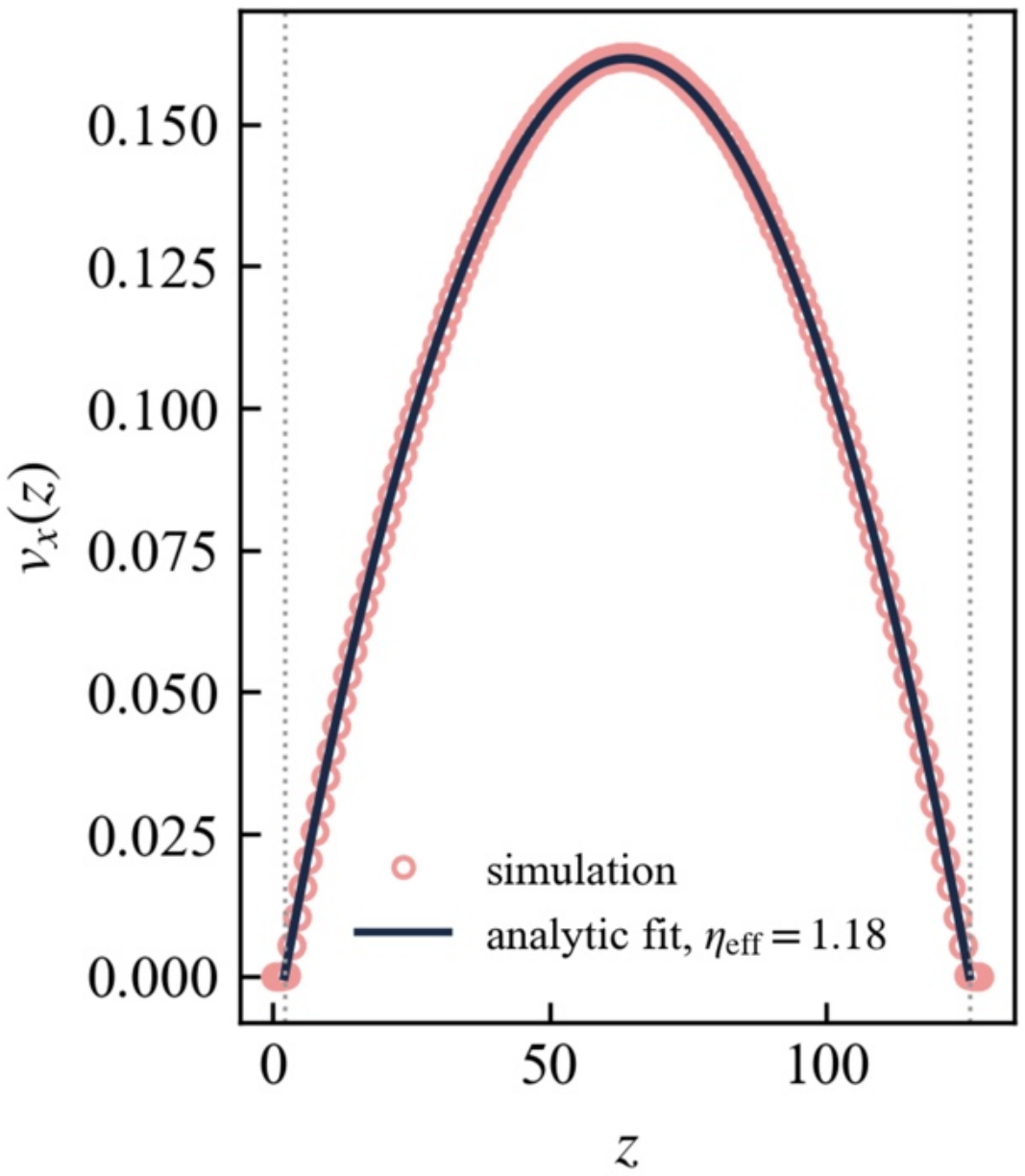

**Extended Data Figure 1 | Validation of the no-slip fluid boundary.** Numerical pure-solvent Poiseuille profile under body force $G = 10^{-4}$, compared with the analytical parabola. The numerical velocity approaches zero at both effective walls. A free-wall fit gives $\eta_{\mathrm{eff}} \simeq 1.18$ and $R^2 = 0.9993$.

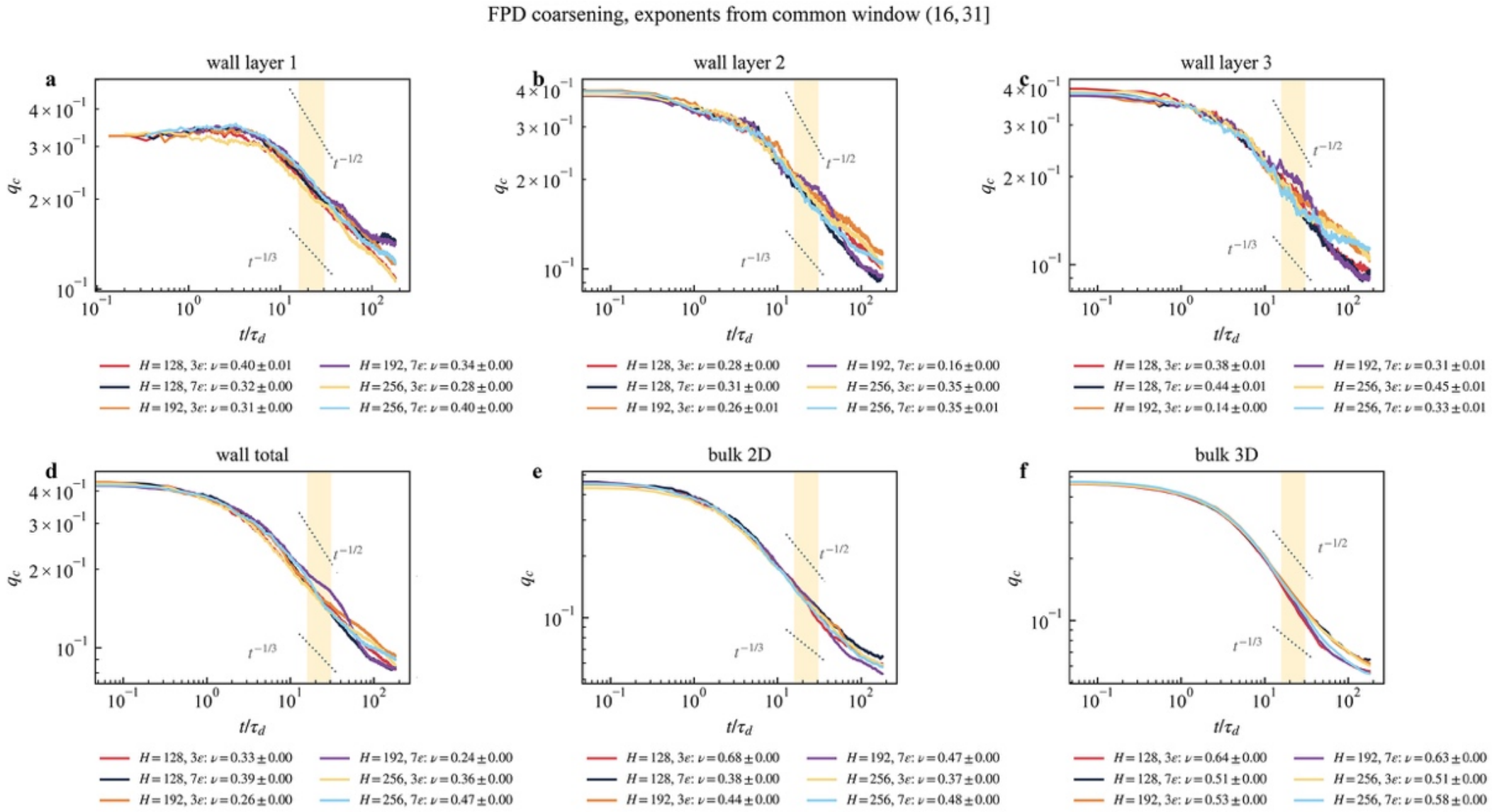

**Extended Data Figure 2 | Complete FPD coarsening data.** Central, lower-wall and upper-wall $q_{c,\parallel}(t)$ curves for all channel heights and wall attractions, with reference slopes, the conservative fitting interval and model-specific robustness intervals.

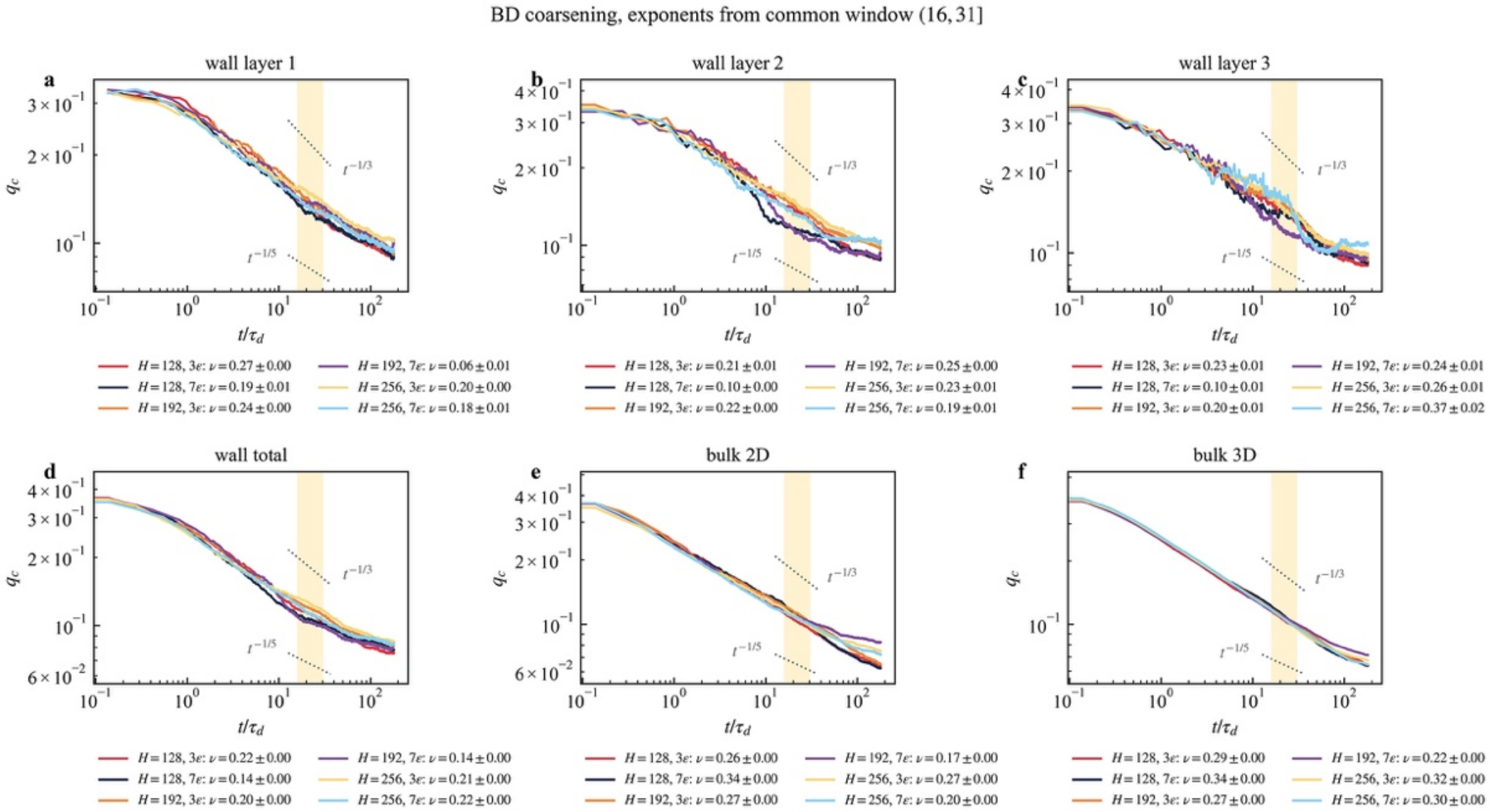

**Extended Data Figure 3 | Complete BD coarsening and running-exponent data.** Free-draining central and wall curves for all channel heights and wall attractions. The central running exponents exhibit extended plateaux, whereas the near-wall values remain time dependent.

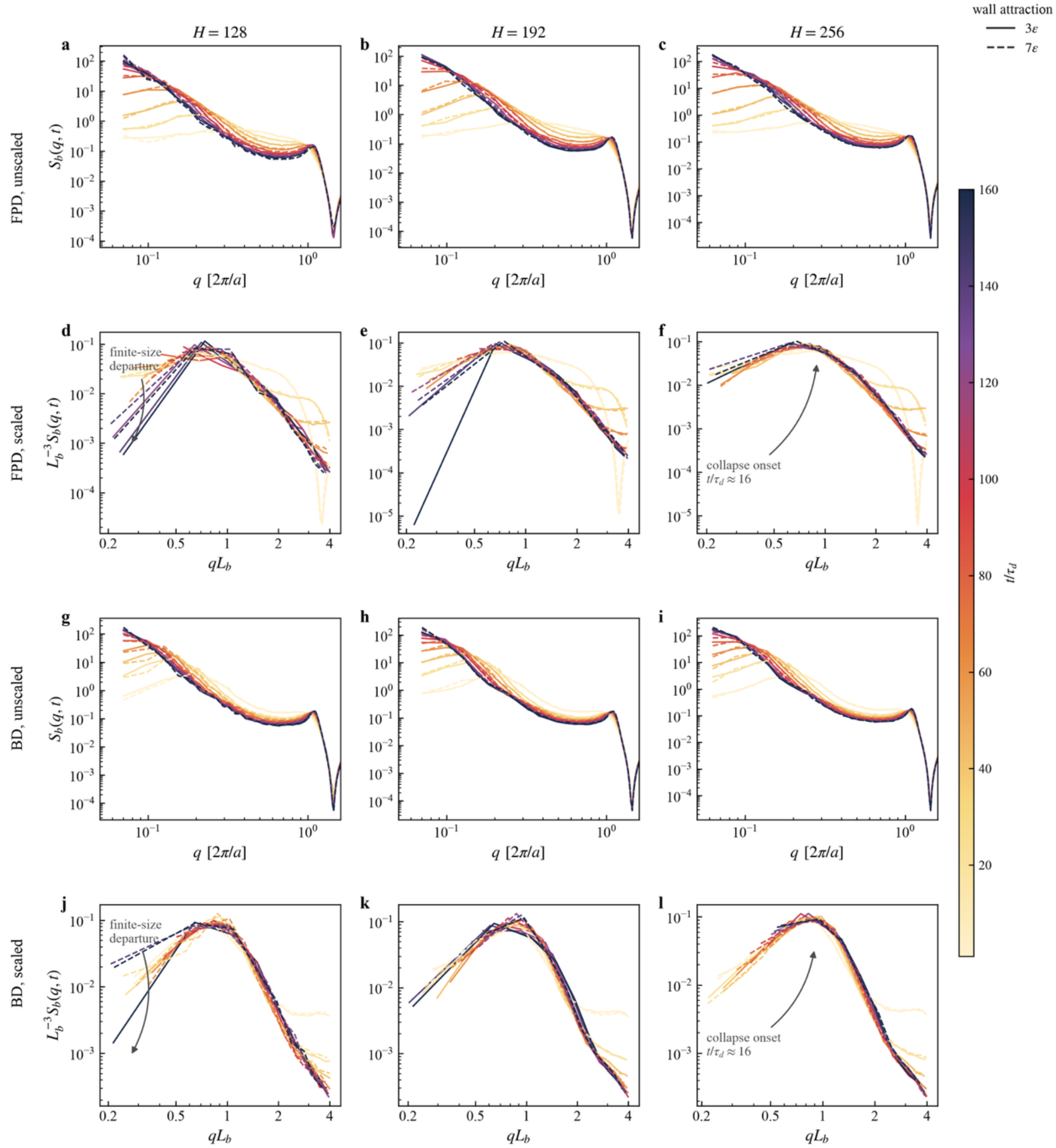

**Extended Data Figure 4 | Three-dimensional structure-factor scaling for all central regions.** Unscaled and scaled central structure factors for all FPD and BD parameter combinations. Collapse onsets, late-time residuals and finite-size departures are indicated. Small-channel data show earlier deviations than the largest-channel data.

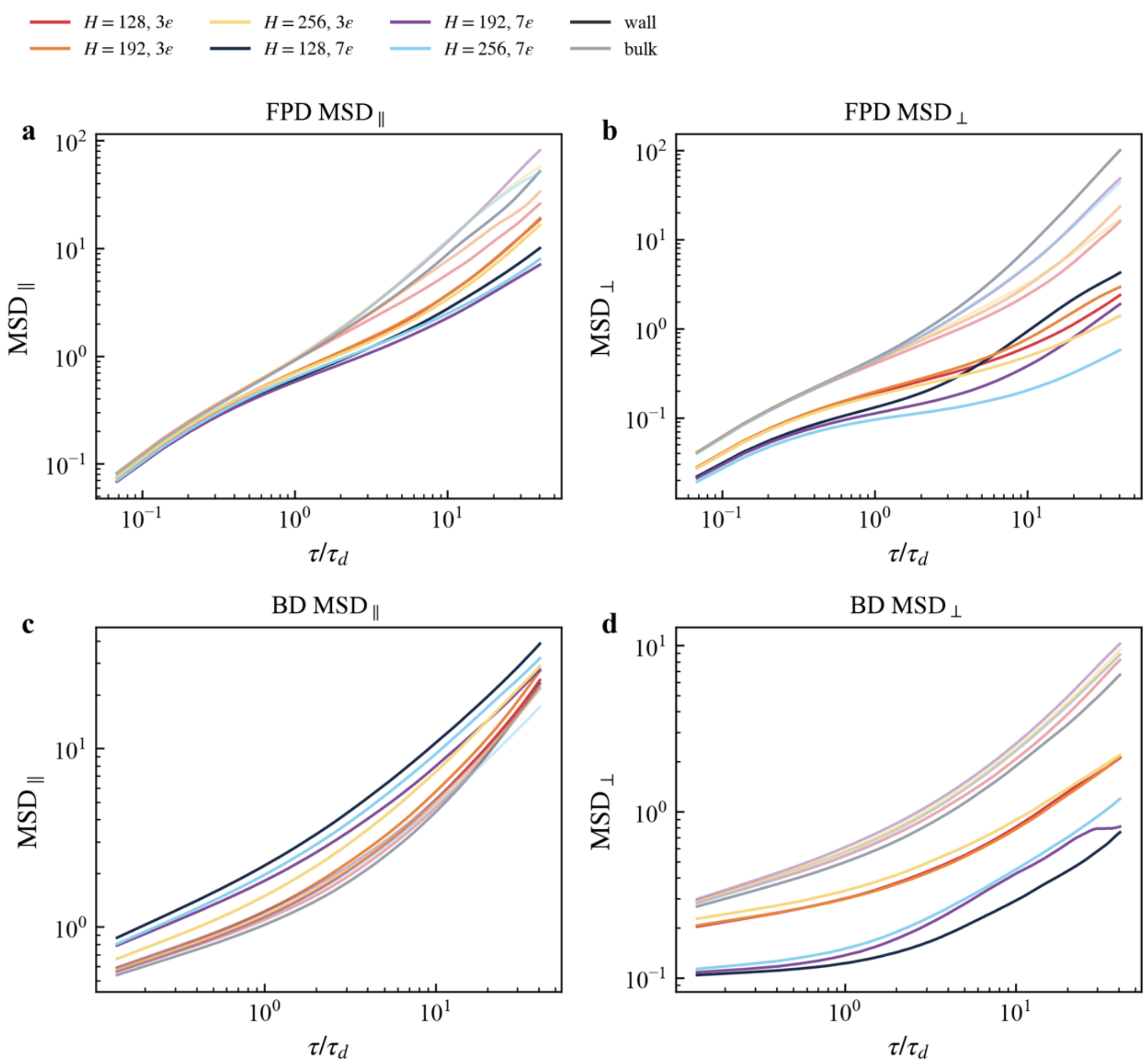

**Extended Data Figure 5 | Complete mean-squared-displacement data.** Parallel and perpendicular particle displacements for all FPD and BD parameter combinations, including $T^* = 0$ where available.

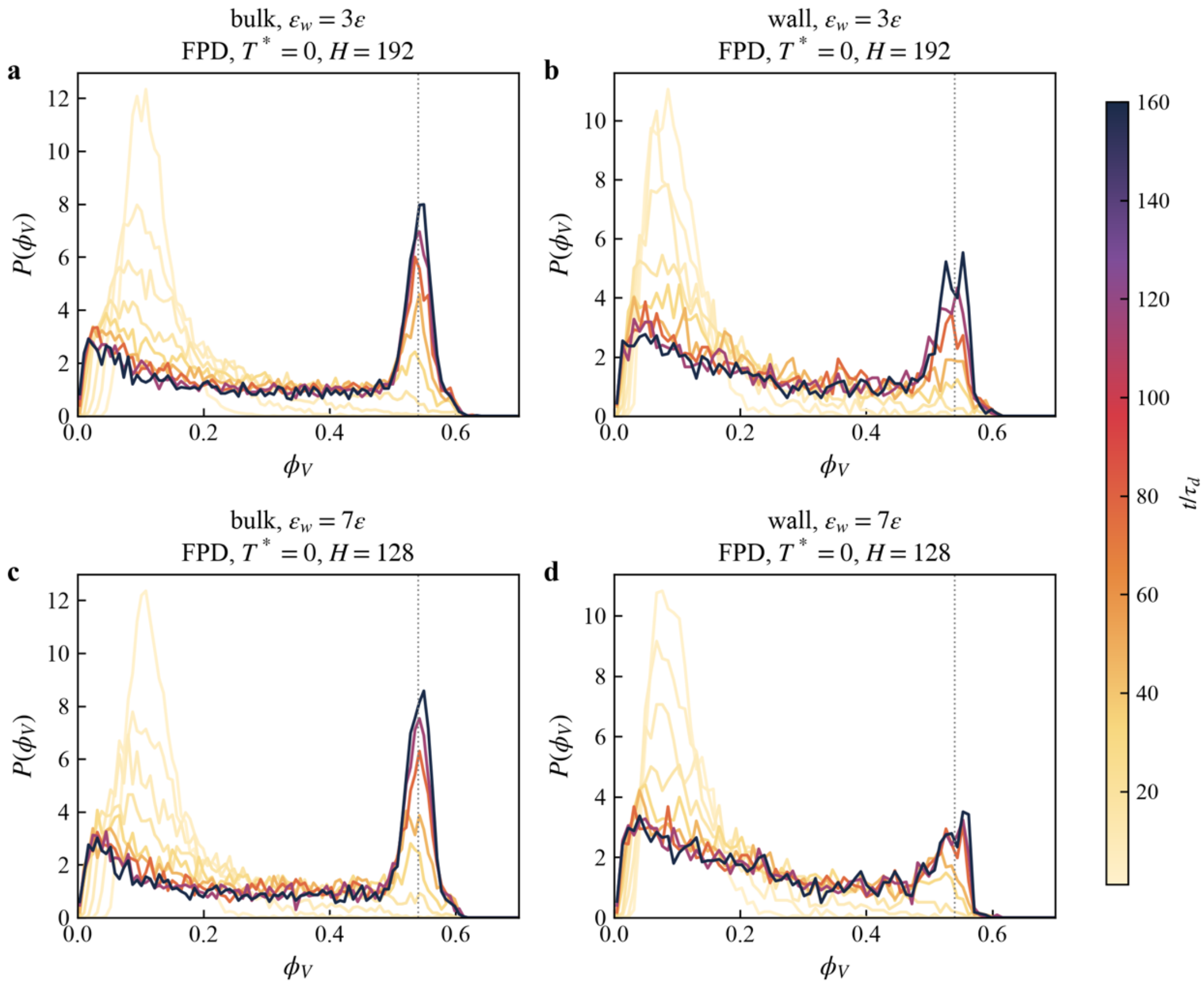

**Extended Data Figure 6 |** Voronoi local-volume-fraction distributions for FPD at $T^*$=0 (H=192, $\varepsilon_w$=3ε; H=128, $\varepsilon_w$=7ε), wall and central regions separately, at nine times $t/\tau_d$ = 1.1–160.

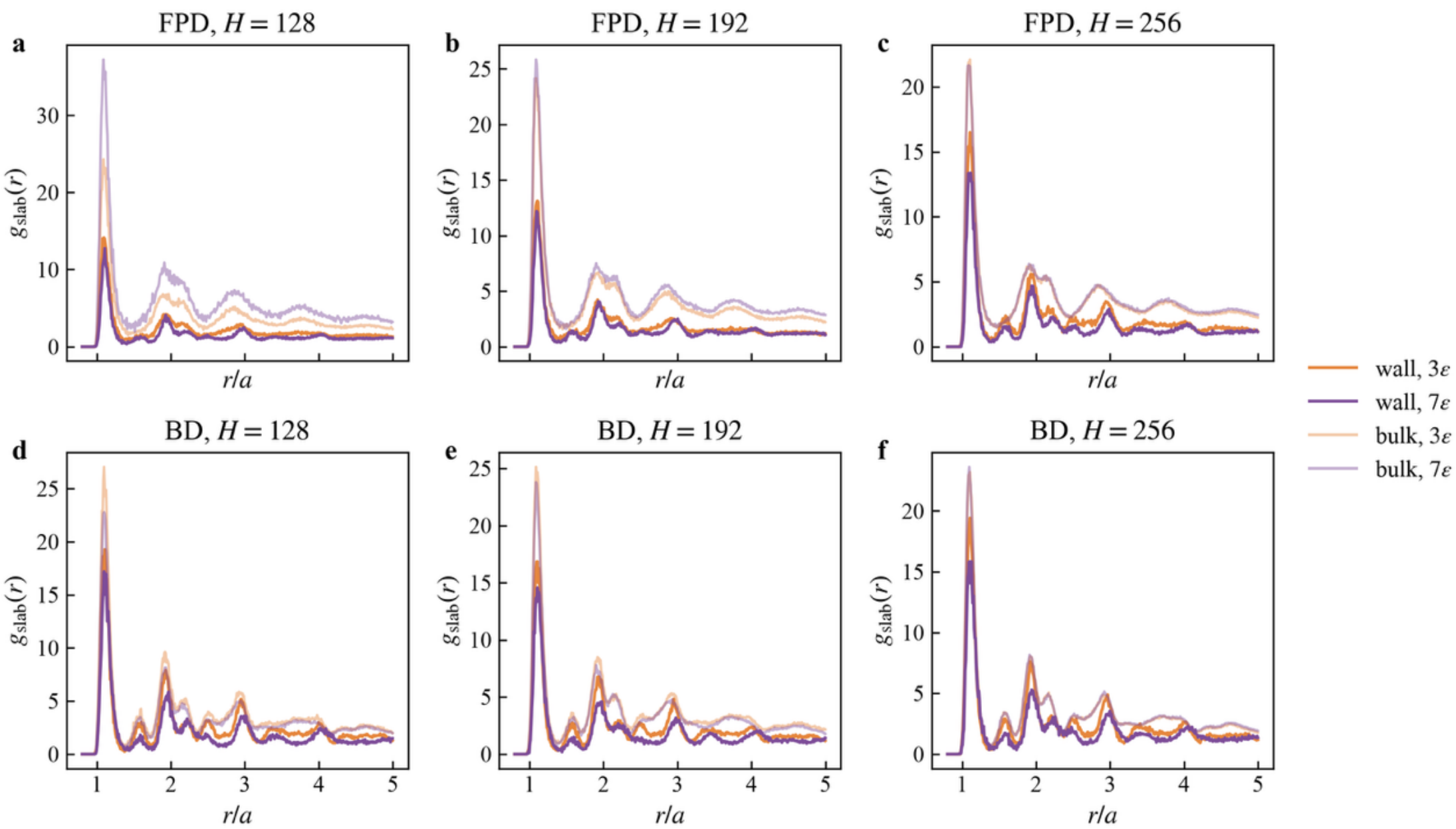

**Extended Data Figure 7 |** Slab radial distribution functions and Voronoi local-volume-fraction distributions for all available combinations of dynamical model, channel height and wall attraction.

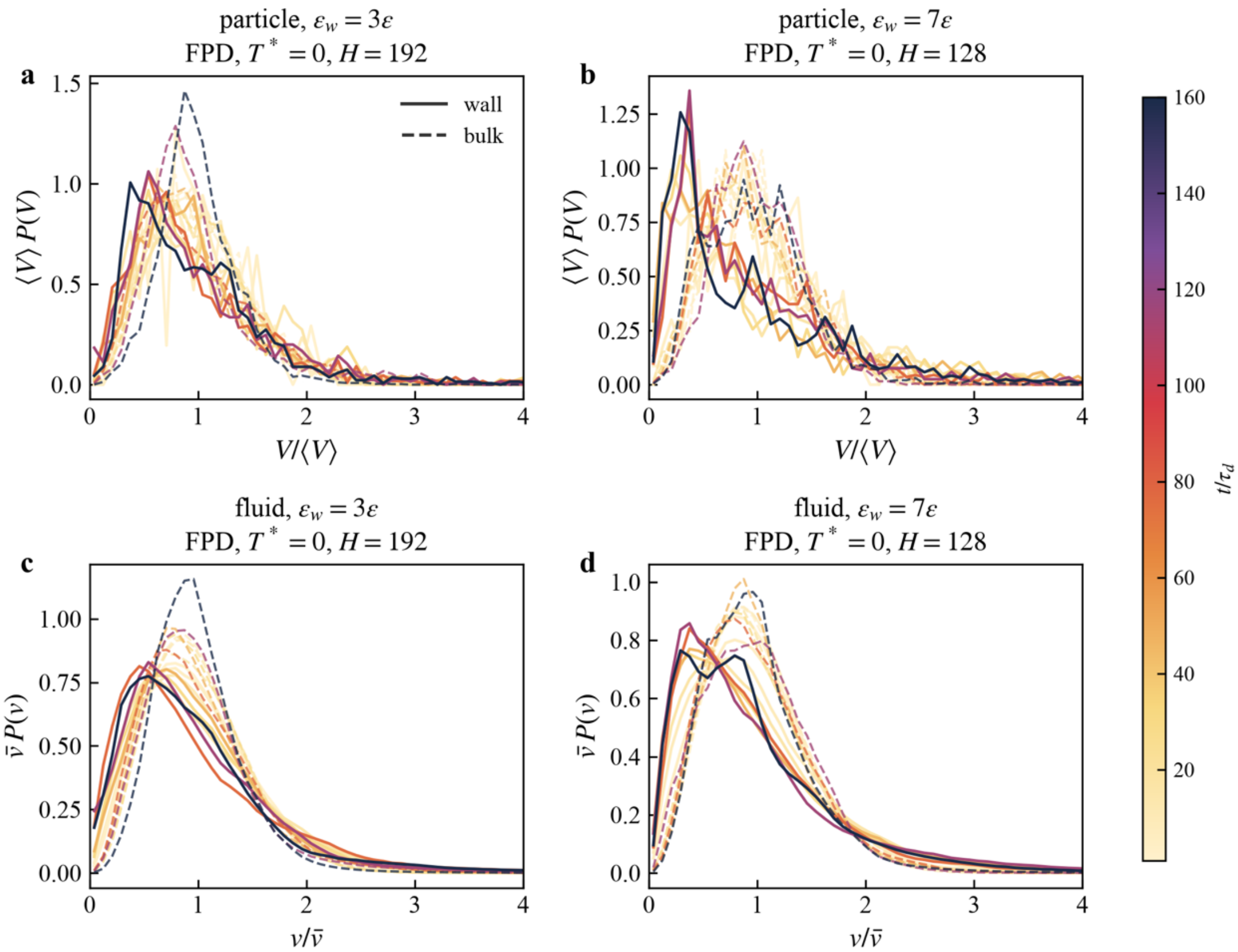

**Extended Data Figure 8 | Particle and fluid velocity distributions.** Scaled particle distributions $\langle V \rangle P(V)$versus $V/\langle V \rangle$for colloid-rich skeleton particles, and scaled fluid distributions $\bar{v}P(v)$versus $v/\bar{v}$. The dominant central portions collapse approximately, whereas high-speed tails and near-wall curves are more sensitive to intermittent restructuring and finite statistics.

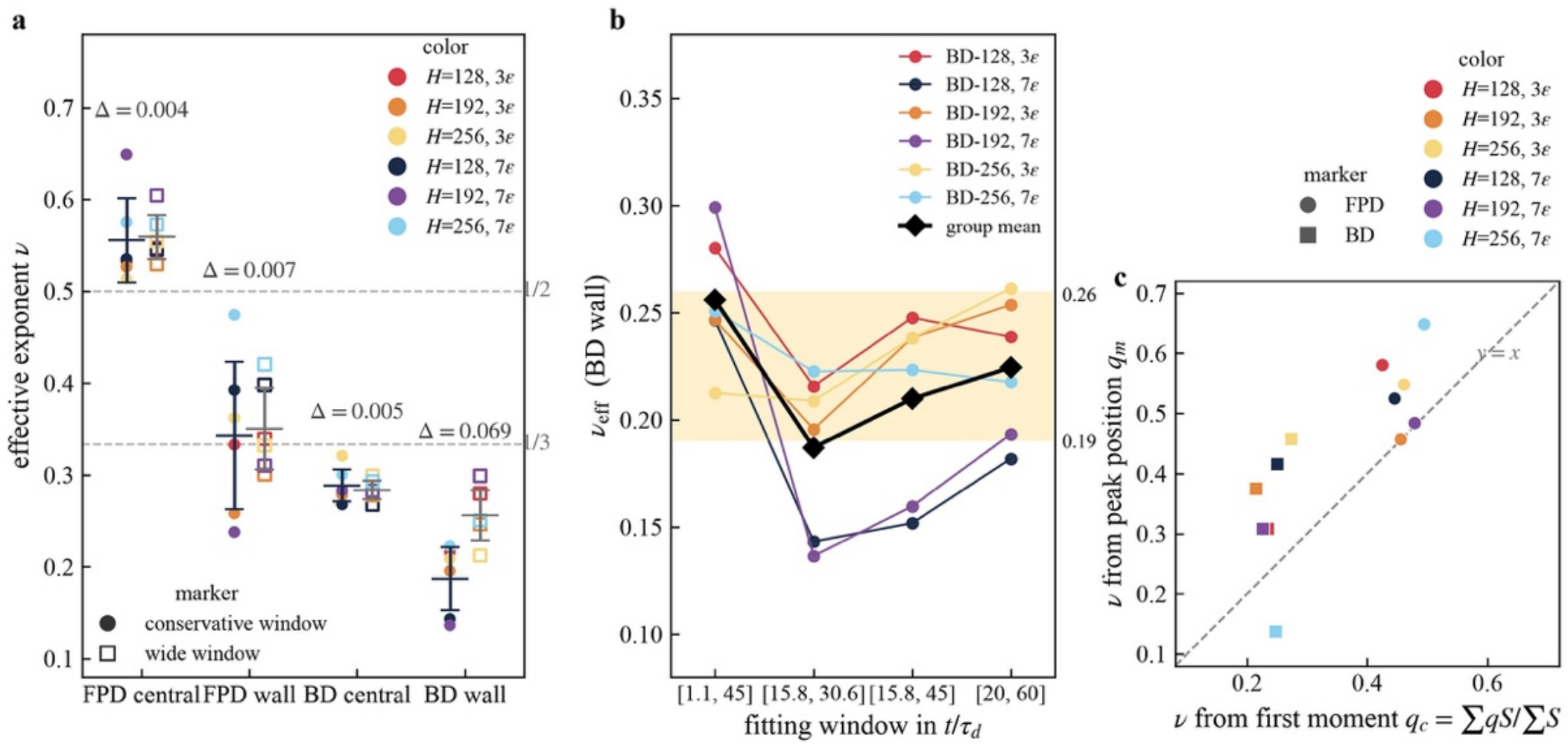

**Extended Data Figure 9 | Fitting-window and characteristic-length robustness.** Comparison of effective exponents obtained from the common conservative interval and wider model-specific intervals. FPD central, FPD wall and BD central exponents vary by no more than 0.007, whereas the BD wall effective exponent spans approximately 0.19–0.26. Results obtained from the first moment and alternative structure-factor length definitions are compared where available.